\documentclass[iop]{emulateapj}

\shorttitle{The Problem of  Type II Migration}
\shortauthors{Hasegawa \& Ida}

\begin{document}

\title{Do Giant Planets Survive Type II Migration?}

\author{Yasuhiro Hasegawa\altaffilmark{1}}
\affil{Institute of Astronomy and Astrophysics, Academia Sinica (ASIAA), Taipei 10641, Taiwan}
\email{yasu@asiaa.sinica.edu.tw}

\author{Shigeru Ida}
\affil{Earth-Life Science Institute, 
Tokyo Institute of Technology, Ookayama, Meguro-ku, Tokyo, 152-8551, Japan}
\email{ida@geo.titech.ac.jp}

\altaffiltext{1}{EACOA fellow}

\begin{abstract}

Planetary migration is one of the most serious problems to systematically understand the observations of exoplanets.  
We clarify that the theoretically predicted type II migration is too fast, as well as type I migration, by developing 
detailed analytical arguments in which the timescale of type II migration is compared with the disk lifetime. 
In the disk-dominated regime, the type II migration timescale is characterized by a {\it local} viscous diffusion timescale, 
while the disk lifetime characterized by a {\it global} diffusion timescale that is much longer than the local one.
Even in the planet-dominated regime where the inertia of the planet mass reduces the migration speed, 
the timescale is still shorter than the disk lifetime except in the final disk evolution stage where the total disk mass decays 
below the planet mass. This suggests that most giant planets plunge into the central stars within the disk lifetime, 
and it contradicts the exoplanet observations that gas giants are piled up at $r \ga 1$AU.
We examine additional processes that may arise in protoplanetary disks: dead zones, photoevaporation of gas,
and gas flow across a gap formed by a type II migrator. Although they make the type II migration timescale
closer to the disk lifetime, we show that none of them can act as an effective barrier for rapid type II migration
with the current knowledge of these processes. We point out that gas flow across a gap and the fraction of the flow
accreted onto the planets are uncertain and they may have a potential to solve the problem. 
Much more detailed investigation for each process may be needed to explain the observed distribution of gas giants 
in extrasolar planetary systems.
\end{abstract}


\section{Introduction} \label{intro}

The accumulation of observed exoplanets around solar-type stars has revealed a number of interesting features of the exoplanets \citep[e.g.,][]{us07}. 
For instance, radial velocity observations suggest that gas giants tend to orbit around their host stars more frequently at $r \ga 1$ AU \citep[e.g.,][]{mml11}. 
As another example, both radial and {\it Kepler} transit observations infer that low-mass planets, also known as super-Earths, are more common than 
massive planets \citep[e.g.,][]{hmj10,bkb11}. These features are prominent in the mass-period diagram and are utilized for examining the current theories of 
planet formation \citep{il04i,mab09}.

Planetary migration is one of the most important processes for interpreting the observations, and hence for understanding planet formation in 
protoplanetary disks \citep[for a most recent review]{kn12}. It arises from tidal interactions between protoplanets and the surrounding gaseous disks 
\citep{gt80,w86,ttw02}. There are two modes in migration that are distinguished by planetary mass: type I and type II migration. In principle, type I migration 
is effective for low-mass planets such as terrestrial planets or cores of gas giants and is well known as one of the most serious problems in 
theories of planet formation. The most advanced studies show that the timescale of type I migration is very rapid ($\sim 10^5$ yr for an earth-mass planet 
at $\sim 1$AU) and its direction is highly sensitive to the properties of disks such as the surface density, temperature, viscosity and opacity of the disk 
\citep[e.g.,][]{pbck09}. Such kinds of complexity in type I migration prevents us from systematically understanding how planets form in protoplanetary disks 
under the action of type I migration. Currently, a number of mechanisms have been proposed for resolving the problem of type I migration. 
For example, \citet{hp11} have recently focused on some kinds of inhomogeneities that are considered to be present in protoplanetary disk and 
investigated how the disk inhomogenties give rise to trapping sites in disks at which the net torque becomes zero and type I migration is halted 
\citep[also see][]{mpt07,il08v,lpm10,hp10,kl12}. The sites are often referred to as planet traps \citep{mmcf06}.\footnote{Since massive bodies that undergo 
type I migration accumulate around planet traps, it is considered that the formation of planetary cores is significantly enhanced there \citep[e.g.,][]{sld11}. 
This invokes another terminology - planet traps are referred to as "convergence zones" in the context of planetary growth \citep{mda11,crp13}. 
It is emphasized that these two are essentially identical with each other.} It is important that planet traps are very useful for systematically examining 
the formation of planetary system architectures \citep{hp11}. 

Type I migration becomes faster as planetary mass increases. However, when a planet becomes massive enough to open up a gap in the disk, 
the planet starts migrating with disk gas accretion, which is called "type II migration" \citep[e.g.,][]{lp86ii,lp86iii,npmk00}. Since the problem of type I migration 
has attracted huge attention and the transition to type II generally makes the migration slower \citep{ward97}, problems of type II migration may have 
been overlooked. As shown below, however, type II migration is {\it not} so slow as anticipated so far. This is because the migration timescale is determined 
by a {\it local} viscous diffusion timescale $\tau_{vis}(r)$, which is generally much shorter than disk lifetime $\tau_{disk}$ as shown in Section \ref{vis}. Since 
the disk lifetime represents a timescale of when gas disks dissipate globally,  $\tau_{disk}$ can be referred to as a {\it global} viscous diffusion timescale. 
Indeed, the population synthesis calculations show that most of formed gas giants migrate to the proximity of the host star \citep[e.g.,][]{il08v}.\footnote{In 
the population synthesis calculations by \citet{mab09,mabn09}, such a problem was not recognized. This is because of their prescriptions for gas accretion 
onto planets. Adopting the prescriptions, the effect of planetary inertia is so enhanced and it halts the migration. For details, see discussions in 
Section \ref{gas_flow}.} This is inconsistent with the radial velocity observations which show that most observed gas giants are piled up at $\ga 1$ AU with 
few population of hot Jupiters \citep{mml11}. 

In this paper, we will make clear the problem of too fast type II migration through analytical arguments. One may wonder the problem of type II migration 
would be solved when a number of additional physical processes are considered such as dead zones, photoevaporation, and gas flow across a gap formed 
by a type II migrator. We will however clarify below that dead zones and photoevaporation do not become crucial for solving the problem of type II migration. While gas flow across a gap formed by a massive planet may have a potential to solve the problem, the details of this process have not been made clear yet.

The plan of this paper is as follows. In Section \ref{vis}, we define two important timescales that are determined for disks regulated by viscous diffusion: 
$\tau_{vis}$ and $\tau_{disk}$,  and then we demonstrate how fast type II migration is by focusing a timescale argument. In Section \ref{disc}, 
we examine three additional physical processes that are often considered in more detail calculations: dead zones, photoevaporation, and gas flow across 
a gap formed by a planet, and discuss how {\it in}effective they are for resolving the problem of type II migration. We summarize our discussion in 
Section \ref{conc}.

\section{Problem of type II migration} \label{vis}
\subsection{Steady accretion disk}

The radial velocity ($v_r$) and disk mass accretion rate ($\dot{M}_{vis}$) due to viscous diffusion are given by \citep[for a review]{fkr92}
\begin{eqnarray}
v_r = \frac{3\nu}{2r} \left(1 + \frac{2r}{\Sigma \nu} \frac{\partial (\Sigma \nu)}{\partial r} \right),
\label{eq:vr}\\
\dot{M}_{vis}=2\pi r \Sigma v_r = 3\pi \Sigma \nu \left(1 + \frac{2 r}{\Sigma \nu} \frac{\partial (\Sigma \nu)}{\partial r} \right),
\label{eq:mdot}
\end{eqnarray}
where $r$ is the disk radius, $\Sigma$ is a disk surface density, $\nu=\alpha c_s H$ is the kinematic viscosity, 
$c_s$ and $H$ are the sound speed and the disk scale height at $r$. We adopt the $\alpha$-prescription for quantifying $\nu$ \citep{ss73}. 
We take a positive sign for the inward motion when defining $v_r$. We first assume a standard self-similar solution for viscous diffusion with 
$\nu \propto r$ \citep{lbp74}. At $r \ll r_d$ where $r_d$ is a characteristic disk size beyond which $\Sigma$ exponentially decays (see Figure \ref{fig1}), 
the second terms of equations (\ref{eq:vr}) and (\ref{eq:mdot}) are neglected compared with the first terms, because 
$\partial (\Sigma \nu)/\partial r \sim \Sigma \nu/r_d$ there:
\begin{eqnarray}
v_r \simeq \frac{3\nu}{2r}, \label{eq:vr2}\\
\dot{M}_{vis} \simeq 3\pi \Sigma \nu = 3 \pi \alpha c_s H \Sigma. \label{eq:mdot2}
\end{eqnarray}
Hereafter we assume these relations.

\subsection{Local and global viscous timescales}

We define a local viscous diffusion timescale as $\tau_{vis}=r/v_r$, using the above model. From equation (\ref{eq:vr2}), this timescale reads as
\begin{equation}
\tau_{vis} \simeq \frac{2 r^2}{3\nu}.
\label{eq:tau_vis1}
\end{equation}
Since it is assumed that $\nu \propto r$, corresponding to disks with constant $\alpha$ and with the disk temperature being proportional to $ r^{-1/2}$,
where $\alpha$ is the parameter of the $\alpha$-prescription for viscosity \citep{ss73}, we obtain $\tau_{vis} \propto r$ (for $r \ll r_d$). 
Equation (\ref{eq:tau_vis1}) can also be expressed in terms of $\dot{M}_{vis}$ and $M_d(r)$, where $M_d(r)$ is the total disk mass within $r$. 
This becomes possible by the fact that  $\Sigma \nu$ is constant. As a result, the surface density becomes $\Sigma \propto 1/r$, and $M_d(r)$ is given as
\begin{equation}
M_d(r) = \int^r 2 \pi r \Sigma(r) dr = 2 \pi \Sigma(r) r^2.
\label{eq:Mdr}
\end{equation}
Then, equation (\ref{eq:tau_vis1}) is written as (see equations (\ref{eq:mdot2}) and (\ref{eq:Mdr}))
\begin{equation}
\tau_{vis} \simeq \frac{M_d(r)}{\dot{M}_{vis}}.
\label{eq:vis_2}
\end{equation} 

We can also apply equation (\ref{eq:vis_2}) for characterizing a global viscous diffusion timescale that is calculated as 
\begin{equation}
\tau_{disk} \simeq \frac{M_d(r_d)}{\dot{M}_{vis}},
\label{tau_disk}
\end{equation}
where $\dot{M}_{vis}$ is obtained at $r \ll r_d$ (but not at $r=r_d$). Since $r_d$ is the initial disk size or the disk radius in which most disk masses are contained 
(see Figure \ref{fig1}), we can call $\tau_{disk}$ as the disk lifetime.

The observations of protoplanetary disks in (sub)millimeter wavelength infer that the median value of $M_d(r_d)$ is $\sim 10^{-2} M_{\odot}$ for 
Classical T Tauri stars (CTTSs) (e.g., \citealt{aw05}; \citealt{aw07}, also see \citealt{wc11} for a most recent review). In addition, the observations of 
the disk accretion rate show that $\dot{M}_{vis} \sim 10^{-8} M_{\odot}$ yr$^{-1}$  for CTTSs \citep[e.g.,][]{hcg98,cmb04}. These observations are consistent 
with the NIR observations which show that $\tau_{disk}$ is a few million years \citep[e.g.,][]{heg95,hll01}. Combining such observations with the relation that 
$\tau_{vis} \propto r$, the local viscous timescale can be written as (see equations (\ref{eq:vis_2}) and (\ref{tau_disk}))
\begin{equation}
\tau_{vis} = \frac{r}{r_d} \tau_{disk} \sim \frac{r}{r_d} \times (10^6-10^7) \mbox{yrs}.
\label{disk_lt}
\end{equation}
Thus, the viscous diffusion timescale (either local or global) can be determined by the characteristic disk radius or by the total disk mass that is contained in the 
characteristic disk radius. 

\subsection{Too fast type II migration} \label{typeII}

We show that a timescale of type II migration is expected to be too short to retain gas giants at large $r$. There are two regimes in type II migration: 
disk-dominated and planet-dominated migration. These regimes are distinguished by the ratio of planetary mass ($M_p$) to the total disk mass 
within an orbital radius ($r_p$) of a gas giant. When 
\begin{equation}
M_p < M_d(r_p) = 2 \pi \Sigma(r_p) r_p^2,
\label{crit_mass}
\end{equation} 
the gas giant migrates with (unimpeded) disk accretion, which we refer to as "disk-dominated" type II migration. 
In this case, the migration timescale is the same as the {\it local} viscous diffusion timescale:
\begin{equation}
\tau_{mig,d}= \frac{M_d(r_p)}{\dot{M}_{vis}}.
\label{mig_d}
\end{equation}  
where $M_d(r_p)$ is the total disk mass within $r_p$. Obviously, 
\begin{equation}
\tau_{mig,d} \simeq \frac{r_p}{r_d} \tau_{disk} \ll \tau_{disk},
\label{mig_d2}
\end{equation}
because generally $r_p \ll r_d$.

When planetary mass is larger than $M_d(r_p)$, disk accretion from outer regions must push the planet rather than the inner disk.
Then, the migration speed is decreased by the inertia of the planet. This regime is referred to as "planet-dominated" regime, 
In this regime, the type II migration timescale is given as \citep[e.g.,][]{sc95,ipp99,il04i}
\begin{equation}
\tau_{mig,p} \simeq \frac{M_p}{\dot{M}_{vis}}.
\label{mig_p}
\end{equation}
When the planet-dominated type II migration is compared with the disk-dominated one, 
\begin{equation}
\tau_{mig,p} \simeq \frac{M_p}{M_d(r_p)} \frac{M_d(r_p)}{\dot{M}_{vis}}
                  =  \frac{M_p}{M_d(r_p)} \tau_{mig,d},
\label{mig_p1}
\end{equation}
Thus, the type II migration slows down when type II migrators are in the planet-dominated regime ($M_p > M_d(r_p)$). However, since 
\begin{equation}
\tau_{mig,p} \simeq \frac{M_p}{M_d(r_d)}
\frac{M_d(r_d)}{\dot{M}_{vis}}=  \frac{M_p}{M_d(r_d)} \tau_{disk},
\label{mig_p2}
\end{equation}
$\tau_{mig,p}$ is still shorter than $\tau_{disk}$, except the final stage of disk evolution which may be able to eventually achieve the condition that 
$M_p \ga M_d(r_d)$ due to the subsequent disk dispersal and/or the growth of planets.

In summary, the timescale of type II migration is generally shorter than the disk lifetime and it suggests that most formed gas giants plunge into the central stars 
within the disk lifetime. Only an exception is gas giants forming in the end phase of disk evolution. If most of disk gas in the end phase is accreted by a giant, 
the condition that $M_p \ga M_d(r_d)$ can be satisfied for the residual disk gas. However, a very fine tuning is needed in this case for the planet to survive from 
type II migration.\footnote{also see the discussion on planetary inertia in Section \ref{gas_flow}.}

\section{Additional physical processes} \label{disc}

As discussed above, the typical timescale of type II migration (either disk-dominated or planet-dominated) is much shorter than the disk lifetime, as long as
an ordinary viscous disk with constant $\alpha$ is considered. To save a gas giant against type II migration, additional processes that may occur 
in protoplanetary disks must be considered. We discuss three possibilities here: dead zones, photoevaporation, and gas flow across a gap formed by a gas giant. 
We will however show that none of them plays a crucial role for the survival of the gas giant.

\subsection{Dead zones} \label{disc_dz}

It is considered that dead zones are present in the inner region of protoplanetary disks \citep[e.g.,][]{g96}. In such a region, the high column density prevents 
gas from being ionized by X-rays from the central stars and cosmic rays. The suppression of ionization there leads to a poor coupling with magnetic fields 
threading disks. As a result, magnetorotational instabilities (MRIs) become inactive and the region is considered as a "MRI-dead" zone. With the $\alpha$
prescription, the strength of turbulence in the dead zone is about $\alpha_{DZ} \simeq 10^{-3}-10^{-6}$ \citep[e.g.,][]{is05} whereas 
$\alpha_{AZ} \simeq 10^{-1}-10^{-3}$ for the MRI-active region in disks \citep[e.g.,][]{pn03}.

When dead zones are incorporated in disk models, one might consider that the problem of rapid type II migration could be resolved, because the speed of 
type II migration ($v_{r} \sim 3\nu / 2r$) will slow down in the dead zones via reduction of $\alpha$ (recall that $\nu \propto \alpha$). 
As discussed below, however, the presence of dead zones essentially leads to the same conclusion as above - 
only planets formed in the end stage of disk evolution may be able to survive.

\begin{figure*}
\begin{minipage}{17cm}
\begin{center}
\includegraphics[width=8cm]{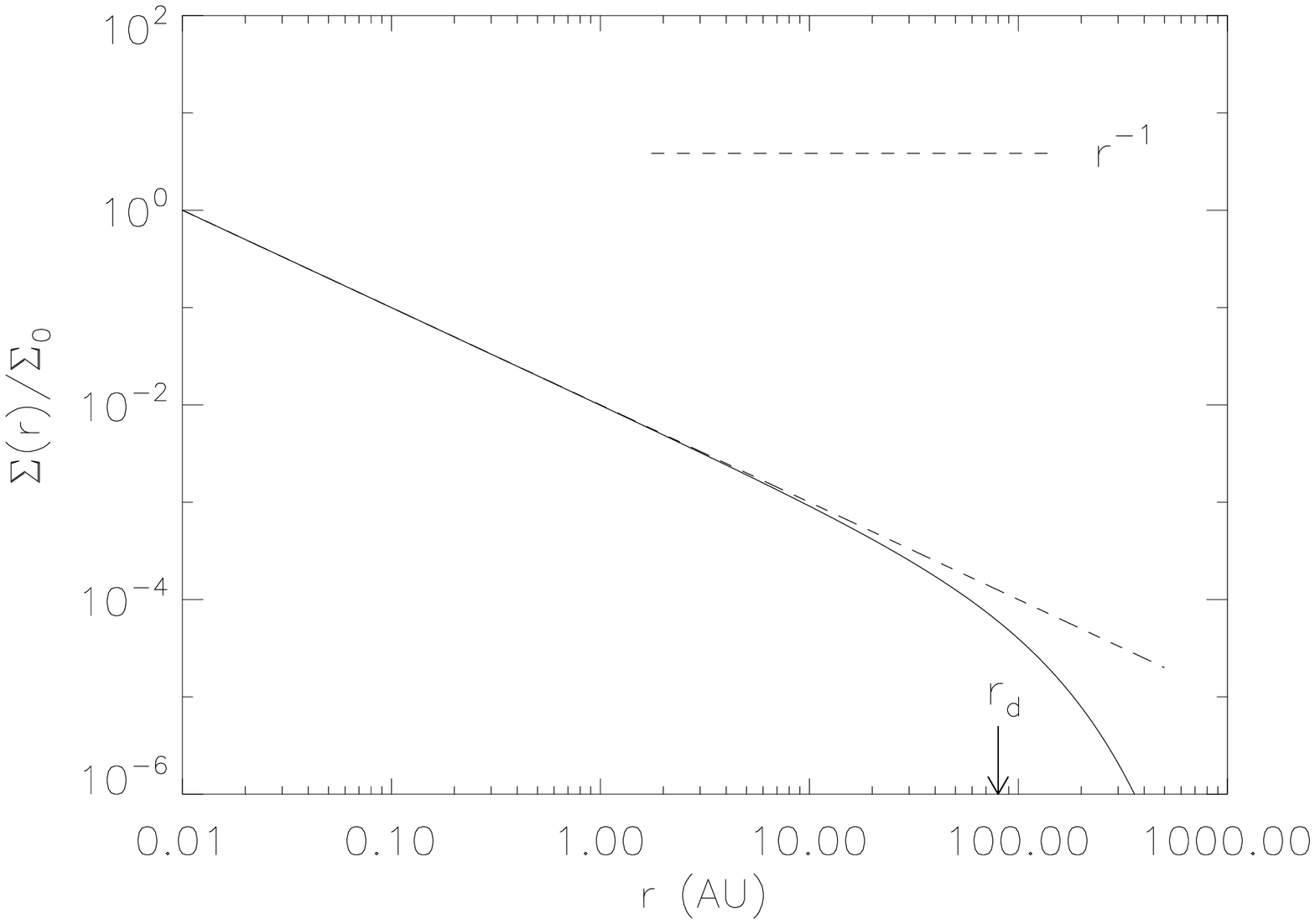}
\includegraphics[width=8cm]{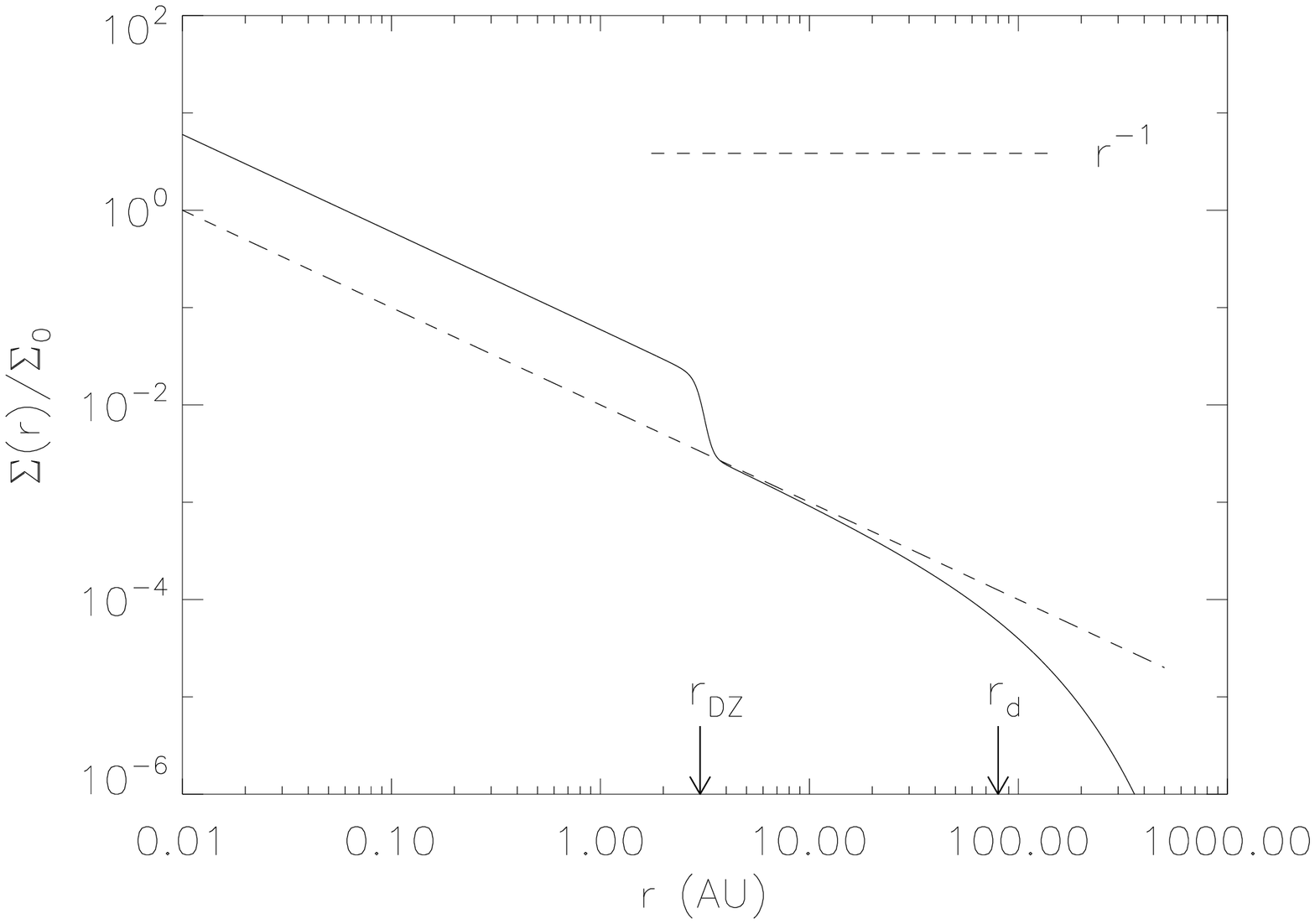}
\caption{Schematic pictures of the surface density of disks and the relevant radial scale. The left panel shows the surface density of a disk and 
the characteristic disk radius ($r_d$) for similarity solutions whereas the right panel shows that the surface density of a disk with a dead zone and 
the characteristic radius of the dead zone ($r_{DZ}$) and the disk size ($r_{d}$). The presence of the dead zone forms a density jump within the 
dead zone.}
\label{fig1}
\end{center}
\end{minipage}
\end{figure*}

We first discuss how the disk structure is affected by dead zones and then examine how the disk timescale is modified by them. 
Figure \ref{fig1} schematically shows that the surface density of a disk jumps up inside a dead zone. 
This occurs because $\dot{M}_{vis} \propto \alpha \Sigma$ (see equation (\ref{eq:mdot2})) and the value of $\alpha$ is much lower in the dead zone. 
In other words, the disk mass is pilled up in the dead zone, because the disk mass that flows from the outer active regions cannot be efficiently transferred there.

We can then define two kinds of lifetimes for disks with dead zones: the lifetimes for "dead" inner disk ($0 < r < r_{DZ}$) and 
"active" outer disk ($r_{DZ} < r < r_d$) (see Figure \ref{fig1}). The former is regulated by the disk mass within the dead zone, 
so that the lifetime of the dead zone is given as 
\begin{equation}
\tau_{disk}^{DZ} = \frac{M_{d}^{DZ}}{\dot{M}_{vis}},
\end{equation} 
where $M_{d,}^{DZ}$ is the disk mass within the dead zone and is written by $M_{d}^{DZ} = M_{d}(r_{DZ})$. 
The lifetime for the active outer disk is determined by 
\begin{equation}
\tau_{disk}^{AZ} = \frac{M_{d}^{AZ}}{\dot{M}_{vis}},
\end{equation} 
where $M_{d}^{AZ}= M_d(r_d)-M_d(r_{DZ})$. As shown in Figure \ref{fig1}, the increment in $\Sigma$ in dead zones can lead to $M_{d}^{DZ}$ that is larger than 
$M_{d}^{AZ}$. Equivalently, $\tau_{disk}^{DZ}$ can become longer than $\tau_{disk}^{AZ}$. As an example, we consider a disk that has $r_d =100$ AU. 
Assuming that $r_{DZ}=10$ AU, $\alpha_{DZ}=10^{-4}$, and $\alpha_{AZ}=10^{-2}$, we find that 
$M_{d}^{AZ} = \int_{r_{DZ}}^{r_d} 2 \pi r \Sigma dr =((r_d - r_{DZ})/r_{DZ})(\alpha_{DZ}/\alpha_{AZ}) M_{d}^{DZ} = 0.09M_{d}^{DZ}$, where
the relation that $M_d \propto \Sigma r \propto r/ \alpha$ has been used. Thus, it is normally expected that $\tau_{disk}^{DZ} > \tau_{disk}^{AZ}$. 

We are now in a position to compare the type II migration timescale with the lifetime of disks with dead zones. Based on the discussion given in the above 
section, the comparison is essentially identical to estimating the mass difference between $M_d^{DZ}$ and either the disk mass within in $r_p$ or 
planetary mass, depending on the regime of type II migration. 

When planets are in the disk-dominated regime ($M_p < M_d(r_p)$), the planets survive only at $r_p \ga r_{DZ}$. This is equivalently $M_d(r_p) \ga M_d^{DZ}$. 
It has been suggested that an outer boundary of the dead zone could be a planet trap for rapid type I migration \citep{mpt09,hp10,hp11}. Recently, 
2D hydrodynamical  simulations have confirmed the robustness of the trap \citep{rsc13}. As a result, gas giants can survive against type II migration 
if they form beyond $r_{DZ}$. It is however more likely to consider that giant planet formation takes place efficiently in dead zones, 
since most disk masses distribute there. Then, planets formed at $r_p < r_{DZ}$ will become hot jupiters due to type II migration. 

For the planet-dominated regime, planets are prevented from falling onto the host stars only when $M_p > M_{d}^{DZ}$. 
This can be achieved only for planets that are formed at the end stage of disk evolution with a condition that $M_p \sim M_d^{DZ}$ being satisfied. 

In summary, while the problem of type II migration becomes less severe by the presence of dead zone, it does not essentially save gas giants from 
significant migration unless planets are formed beyond dead zones. More specifically, we obtain that the survival condition is that $r_p\sim r_{DZ}$ for disks 
in which most disk masses distribute within $r=r_{DZ}$. For this case, the ratio $\tau_{mig}/\tau_{disk}$ becomes closer to unity by a factor of 
$M_d(r_d)/M_d(r_{DZ}) \sim r_d/r_{DZ}$ than that in the case without dead zone. However, the ratio does not exceed unity for most of gas giants that 
form in the dead zone with the high surface density.

\subsection{Photoevaporation} \label{disc_pe}

Photoevaporation of gas disks may be another process for potentially saving gas giants from plunging into the central stars. However, current understanding of
photoevaporation shows that photoevaporation would not be a key agent to the problem of type II migration, although more detailed observations and simulations
on EUV flux from central stars are required.

\subsubsection{Basic picture}

Photoevaporation arises from heating up gas in protoplanetary disks due to high energy photons emitted from the central and external stars 
\citep[see][for a most recent review]{a11}. It is currently considered as one of the most promising processes for dissipating gas disks 
at the end stage of disk evolution and for determining disk lifetime. In principle, the high energy photons ionize atoms and dissociate molecules in disks and 
result in evaporating them from disk surfaces. It has been recently shown that disk winds could also play a similar role \citep{smi10,bs13}. 

There is a critical disk radius beyond which photoevaporation of gas plays a dominant role in disk evolution. The radius is referred to as 
the gravitational radius and is defined by 
\begin{equation}
r_g=\frac{GM_*}{c_s^2} \sim 10 \mbox{AU} \left( \frac{M_*}{M_{\odot}} \right) \left( \frac{T}{10^4 \mbox{K}} \right)^{-1}.
\label{gra_r}
\end{equation} 
This is derived from comparing the gravitational energy of gas with its thermal energy. It is noted that a more rigorous derivation predicts that 
the gravitational radius is likely to be smaller than that estimated from equation (\ref{gra_r}) by a factor of a few.\footnote{In a more rigorous treatment, 
$r_{g}$ is derived from the Bernoulli's equation, $0= v_K^2/2 + w - GM_*/r_g = (1/(\gamma-1) + 1)(P/\rho) -  GM_*/2r_g = c_s^2/(\gamma-1) -  GM_*/2r_g$ 
where $w=u+P/\rho$ is the specific enthalphy, $u$ is the specific internal energy and $\gamma$ is 5/3 and 7/5 for monoatomic and diatomic molecules. 
Then the gas pressure in the flow plays a counteractive role against the gravitational energy and results in reduction of $r_g$ \citep{sjh93,mjm03,l03}. 
The shrinkage of $r_g$ is confirmed by more detailed simulations \citep[e.g.][]{fmj04,oec10}.} 

Historically, the equation was derived for photoevaporation induced by EUV photons that can ionize hydrogen atoms \citep{sjh93,hjl94}. It is important that 
the value of $r_g$ highly depends on the gas temperature that is heated by high energy photons. For instance, EUV heating results in the gas temperature 
of $10^4$ K whereas FUV and X-ray heating attains $10^2-10^3$ K \citep[e.g.,][]{gh09}. The low temperature by FUV and X-rays increases $r_g$ by about 
a factor of 10 from that for EUV. Thus, only EUV heating can evaporate the inner region of disks.

The recent detailed studies have shown that the effects of photoevaporation on disk evolution are different for different sources 
(the central stars vs the external ones). When photoevaporation by the central star is considered, it can lead to the formation of a gap/inner hole 
in the gas disks \citep{cgs01,mjh03,gdh09,oec10}. This arises from the combination of photoevaporation with viscous diffusion and can be understood 
as what follows. In the case of heating by the central star, photoevaporation is the most efficient around $r_g$ because the flux is proportional to square of 
the distance from the central star \citep[e.g.,][]{mjh03,r04}. Viscous diffusion is fast in inner regions, and when photoevaporation is considered, 
gas is no more supplied from outer regions. As a result, gap formation proceeds, followed by rapid viscous dissipation of the inner disk. 
For the case of heating by the external stars, disks are evaporated in an outside-in manner down to $r_g$ \citep{jhb98,mjh03}. 
This occurs because the outer disks have a larger cross section for photoevaporative photons.

\subsubsection{Slowing down conditions} \label{disc_photoeva0}

\begin{table*}
\begin{minipage}{17cm} 
\begin{center}
\caption{Summary of the slowing-down/survival conditions for photoevaporation}
\label{table1}
\begin{tabular}{ccccc}
\hline 
Source stars    & Gap formation     & Positions of planet formation   & Slowing down conditions                                 & Survival conditions                   \\ \hline 
Central            & No                       &  $r_p \gg r_g$                          & $M_p \ga M_d(r_d)$                                         & Fine tuning                               \\
Central            & No                       &  $r_p < r_g$                             & $M_p \ga M_d(r_d)$                                         & Fine tuning                               \\
Central            & Yes                      &  $r_p < r_g$                             & $M_p \ga M_d(r_d)^1$                                         & Fine tuning                                \\
Central            & Yes                      &  $r_p > r_g$                             & $\dot{M}_{pe} \tau_{disk} \ga M_d(r_d)$            & Gap formation is maintained    \\
                        &                             &                                                 &                                                                           & when planets pass at $r=r_g$   \\ 
External          & No                        &  $r_p < r_g$                             & Gas disposal quickly proceeds at $r=r_p$       & Fine tuning                                  \\
                       &                              &                                                 & before planets migrate towards the central stars  &                                                \\
\hline
\end{tabular}
$^1$ This is applicable only if planet formation completes just after a gap is formed. If planet formation proceeds well after the gap formation and 
gaps are maintained for the rest of the disk lifetime, formed planets cannot grow to gas giants.
\end{center}
\end{minipage}
\end{table*}

We now discuss disk evolution affected by photoevaporation and its effect on type II migration. 
Here we summarize slowing-down conditions that can be achieved by photoevaporation (see Table \ref{table1}).
We mainly consider photoevaporation from the central star. We define $\dot{M}_{pe}$ by the total photoevaporation rate integrated over the disk. 
For simplicity, we approximate that $\dot{M}_{pe}$ is contributed only from the region at $r \sim r_g$. 
We also define $\dot{M}_{vis}$ by disk mass accretion rate due to viscous diffusion at $r_g \ll r \ll r_d$, which is not affected by photoevaporation.
We first focus on the case that no gap in a gas disk is formed by photoevaporation.

When planet formation proceeds beyond the gravitational radius ($r_p \gg r_g$), the migration rate is the same as that in the case without 
the effect of photoevaporation, so that we can apply the same argument as done in Section \ref{typeII}. For the case that $r_p < r_g$, on the other hand, 
the disk accretion that exerts torque onto the planet and contributes to type II migration is decreased from $\dot{M}_{vis}$, while global disk depletion still 
occurs with $\dot{M}_{vis}$. It is therefore expected that survival chance of giant planets can increase according to photoevaporation. We thus examine 
the case that $r_p < r_g$ in detail. In this case, the net disk accretion rate 
$\dot{M}_{acc}$ is given as \citep[e.g.,][]{r04}
\begin{equation}
\dot{M}_{acc} = \dot{M}_{vis} - \dot{M}_{pe}.
\label{M_acc}  
\end{equation}
Since it is currently considered that no gap formation proceeds, $\dot{M}_{vis} > \dot{M}_{pe}$.
Then type II migration timescales, equations (\ref{mig_d}) and (\ref{mig_p}), are written as
\begin{eqnarray}
\tau_{mig,d}=\frac{M_d(r_p)}{\dot{M}_{acc}} & \mbox{ [for the disk-dominated regime]}, \\
\label{eq:tmigdpe}
\tau_{mig,p}=\frac{M_p}{\dot{M}_{acc}} & \mbox{ [for the planet-dominated regime]}.
\label{eq:tmigppe}
\end{eqnarray}
Substituting $\dot{M}_{acc} =f_{pe} \dot{M}_{vis}$ with $f_{pe}=1 - \dot{M}_{pe}/ \dot{M}_{vis}$ ($0 \le f_{pe} < 1$), the above two equations become
\begin{equation}
\tau_{mig,d}=\frac{M_d(r_p)}{f_{pe} M_d(r_d)} \frac{M_d(r_d)}{ \dot{M}_{vis}}  =\frac{M_d(r_p)}{f_{pe} M_d(r_d)} \tau_{disk} 
\label{eq:tmigdpe1}
\end{equation}
and
\begin{equation}
\tau_{mig,p}=\frac{M_p}{f_{pe} M_d(r_d)}  \frac{M_d(r_d)}{\dot{M}_{vis}} =\frac{M_p}{f_{pe} M_d(r_d)}  \tau_{disk},
\label{eq:tmigppe1}
\end{equation}
where we have used $\tau_{disk} = M_d(r_d)/\dot{M}_{vis}$ because the global disk depletion timescale is not affected by photoevaporation \citep{r04}.
Thus, when photoevaporation is incorporated in disk models, the type II migration slows down by a factor of $1/f_{pe}$ (since $f_{pe} < 1$), 
even if gap formation does not proceed.

We now discuss more crucial slowing-down conditions that are needed for saving gas giants against rapid type II migration. Based on equations 
(\ref{eq:tmigdpe1}) and (\ref{eq:tmigppe1}), the condition is $f_{pe} < \max[M_d(r_p), M_p]/M_d(r_d)$. More specifically, the survival condition, 
$\tau_{mig}/\tau_{disk} > 1$, is satisfied either if $f_{pe} \ll 1$ or if $M_d(r_d) \la M_p$. The former case is equivalent to $\dot{M}_{pe} \ga \dot{M}_{vis}$, 
which implies formation of a gap by photoevaporation. We therefore neglect the case here (see the discussion below). For the latter case, 
a modest value of $f_{pe}$ ($\dot{M}_{vis} > \dot{M}_{pe}$) is allowed. Thus, the more crucial slowing-down condition for the case of no gap formation is 
$M_d(r_d) \la M_p$.

We consider the case that photoevaporation induces the formation of a gap in the gas disk. Although the self-similar solution cannot be applied to 
the gap formation case, the arguments developed in the above sections may be able to be applied to this case as well except detailed numerical factors. 
It is noted however that we need to consider the lifetime of the inner disk separately from the lifetime of the outer disk that would be similar to 
the global disk lifetime ($\tau_{disk} = M_d(r_d)/\dot{M}_{vis}$). This is because the gap formed at $r=r_g$ divides the disk into the outer and inner ones.

For planets formed in the inner disk ($r_p < r_g$), the same arguments as planets in the dead zone are applied by replacing $r_{DZ}$ and 
$M_d^{DZ}$ with $r_d$ and $M_d(r_g)$ (see Section \ref{disc_dz}): Planets can undergo slowed-down type II migration only at $r_p \sim r_g$ or 
when $M_d (r_g) \la M_p$. Since it is currently considered that planet formation takes place in the inner disks, the most likely condition for slowing down 
type II migration is $M_d(r_g) \la M_p$. We note that this argument can be applied only if gas giants are formed just after a gap is opened 
by photoevaporation. If the formation of planets takes place well after the gap formation and the gap is maintained for the remaining disk lifetime, 
then such planets cannot grow to gas giants. This is because the disk lifetime of the inner disk is much shorter than $\tau_{disk}$ ($r_g \ll r_d$), 
so that gas in the inner disk is rapidly accreted onto the central star.

When a planet is formed in the outer disk ($r_p > r_g$), a gap formed at $r_g$ can act as a barrier for subsequent type II migration 
if the gap is opened before it passes $r_g$. For this case, the slowing down condition is essentially identical to the gap opening condition 
that is given as $\dot{M}_{vis} \la \dot{M}_{pe}$ ($f_{pe} \ll 1$), or $M_d(r_d) \la \dot{M}_{pe} \tau_{disk}$.

We have so far discussed photoevaporation from high energy photons from the central star. If we consider external stars, the disk is removed only in outer parts 
\citep[e.g.,][]{jhb98,mjh03}. It decreases $\tau_{disk}$ due to the reduction in $r_d$ (see equation (\ref{tau_disk})). Nonetheless, type II migration significantly 
slows down only if the disk gas at $\sim r_p$ is quickly removed after the formation of a planet and before it migrates to the proximity of the central star.
It requires a fine tuning, so that most of planets are not saved.

\subsubsection{Survival conditions} \label{disc_photoeva}

We now discuss how the slowing-down conditions derived in the above section can be feasible in protoplanetary disks, in the case of internal photoevaporation.
Table \ref{table1} summarizes such survival conditions. For the case that the slowing down condition is that $M_p \ga M_d(r_d)$, 
a fine tuning is needed for saving gas giants against type II migration, regardless of whether or not gap formation proceeds due to photoevaporation. 
This was already discussed in Sections \ref{typeII} and \ref{disc_dz}. \citet{all02} also derived the same conclusion.

The most promising situation may be that the slowing down condition is $M_d(r_d) \la \dot{M}_{pe} \tau_{disk}$ (see Table \ref{table1}). 
This is because type II migrators can be halted at $r=r_g$ due to the formation of a gap by photoevaporation if the gap is kept open when planets pass it. 
In fact, this scenario is already suggested by \citet{mjm03}. Nonetheless, this case may not be favored in protoplanetary disks. 
First, the above slowing down condition becomes $\dot{M}_{pe} \ga 10^{-8}$ M$_{\odot}$ yr$^{-1}$, 
when $M_d(r_d) \sim \mbox{a few} \times 10^{-2}M_{\odot}$ and $\tau_{disk} \sim \mbox{a few} \times 10^{6}$ yr, 
both of which are typical values for disks around CTTSs \citep{wc11}, are adopted. Although the recent models show that the photoevaporation rates by 
X-ray or FUV can afford such a high rate (Table \ref{table2}), the rate may cause another problem:  
when FUV- and X-ray-induced photoevaporation is considered, $r_g$ expands to $\sim 100$AU. The significant reduction in the surface density 
inside $\sim 100$AU, corresponding to $\dot{M}_{pe} \ga 10^{-8}$ M$_{\odot}$ yr$^{-1}$, may inhibit the formation of gas giants that is most efficient around 
1-10 AU \citep[e.g.,][]{il04i}. Second, when EUV opens up a gap at a few AU, it will occur at $\dot{M}_{pe} \ga 10^{-10}$ M$_{\odot}$ yr$^{-1}$ for EUV 
(Table \ref{table2}; also see discussion in Section \ref{disc_photoeva1}). This implies that the gap may not open until the final disk evolution stage of 
$M_d(r_d) <  $ a few $\times10^{-4}M_{\odot}$, which may be too late to save gas giants efficiently.

In summary, it is unlikely that photoevaporation can easily save gas giants against type II migration.

\begin{table*}
\begin{minipage}{17cm} 
\begin{center}
\caption{Summary of the recent results of $\dot{M}_{pe}$}
\label{table2}
\begin{tabular}{cccc}
\hline
References                                                & \citet{fmj04}                 & \citet{oec11}                          & \citet{gdh09}                       \\ \hline 
Flux                                                            & EUV only                     &  EUV + X-rays                        &  EUV + X-rays + FUV         \\
$\dot{M}_{pe}$ (M$_{\odot}$ yr$^{-1}$)    & $\sim 10^{-10}$           & $ \sim 1.4 \times 10^{-8}$      & $ \sim 3 \times 10^{-8}$     \\
\hline
\end{tabular}
\end{center}
\end{minipage}
\end{table*}

\subsubsection{Photoevaporation by EUV} \label{disc_photoeva1}

As discussed above, EUV heating may play the most crucial role in gap formation in the inner region of disks, and hence in saving gas giants against 
type II migration. Based on the above estimate, this becomes possible if $\dot{M}_{pe} \ga 10^{-8}$ M$_{\odot}/{\rm yr}$. Although \citet{fmj04} suggested 
a much smaller value ($\dot{M}_{pe} \sim 10^{-10}$ M$_{\odot}/{\rm yr}$), $\dot{M}_{pe}$ due to EUV heating is still unclear because the emission of EUV 
is readily absorbed by ISM and it is very difficult to determine EUV flux strightfowardedly \citep[references herein]{gdh09}. 

Another issue to be considered is the dependence of EUV flux on mass accretion rates onto the central star ($\dot{M}_{acc}$). 
This is because the dependence is directly related to whether a gap is created or not. It is considered that FUV flux is produced by accretion shock of 
disk gas onto the central star, so that a positive correlation between FUV flux and $\dot{M}_{acc}$ is expected. Actually, the recent observation 
\citep{ich11} suggests that FUV flux is proportional to accretion luminosity ($\propto \dot{M}_{acc}$). With such a correlation, 
FUV heating is difficult to make a gap in the disk. If photoevaporation due to FUV heating is making a gap, 
$\dot{M}_{acc}$ decreases. Accordingly, FUV flux decreases as well, which prevents the accomplishment of the gap opening. 
If EUV flux also has a correlation with $\dot{M}_{acc}$, a gap would not be opened by this self-regulation process.\footnote{\citet{mjh03} numerically 
show that a gap is formed even if EUV heating is directly related to the disk accretion. It is not clear why they found a gap.}

Ultimate energy source for high energy photons from the central star is disk gas accretion. Nonetheless, it is not necessary to assume that 
all the photons should follow the same dependence on $\dot{M}_{acc}$. This is because emission mechanisms can be different for photons with 
different wavelengths. For example, X-rays can be considered to have a weak dependence on $\dot{M}_{acc}$ \citep{def09}. 
Currently, the emission mechanism and the $\dot{M}_{acc}$ dependence have not been clear for EUV. 
They should be clarified as well as the EUV flux amplitude in order to make clear how important is the effect of photoevaporation on type II migration.

\subsection{Gas flow across a gap formed by giant planets} \label{gas_flow}

We have so far discussed physical processes that can affect the disk structure globally. In this section, we focus on a process that takes place locally in disks. 
More specifically, we examine the flow of gas around a gap formed by a massive planet.\footnote{Note that in Section \ref{disc_pe}, 
we have focused on a gap formed by photoevaporation, whereas we here discuss a gap formed by gravitational perturbations from a planet.}

It is well established that type II migration divides disks into the inner and outer ones by forming a gap due to the tidal torque between massive planets and 
the surrounding gaseous disks \citep[e.g.,][]{lp86ii,lp86iii,npmk00}. In the above sections, we have adopted the {\it innate} disk accretion rate 
($\dot{M}_{vis}$ or $\dot{M}_{acc}$) for estimating the timescales of type II migration (see equations (\ref{mig_d}) and (\ref{mig_p})). 
This is essentially identical to assuming that disk accretion from the outer disk {\it fully} contributes to pushing giant planets inward. With this assumption, 
we led to a conclusion that type II migration is too fast for gas giants to survive at large orbital periods. 

We here discuss a possibility that only a fraction of disk accretion from the outer disk can involve moving massive planets. 
This can be achieved if gas in the outer disk flows into the inner disk across a gap and the gas is subsequently accreted onto the central star. 
If this is the case, it is expected that the timescale of type II migration becomes longer than the previous estimate. This occurs because 
the {\it effective} disk accretion rate that regulates the migration timescale reduces (see equations (\ref{mig_d}) and (\ref{mig_p})). 
More physically, the {\it effective} mass of the outer disk that essentially pushes a planet inward becomes smaller due to the gas flow 
from the outer to inner disk across a gap. 

We estimate how type II migration slows down by the gas flow. This can be done by decomposing the innate disk accretion rate ($M_{vis}$), 
that originates from the outer disk and involves pushing a gas giant inward, into the following components:
\begin{equation}
\dot{M}_{vis}= \dot{M}_{cross} + \dot{M}_{p} + \dot{M}_{mig},
\label{decomp}
\end{equation} 
where $\dot{M}_{cross}$ is the component of gas that flows into the inner disk across a gap without both pushing and being accreted 
by the planet, $\dot{M}_{p}$ is that of gas that flows into the gap and is eventually accreted by the planet, and $\dot{M}_{mig}$ is that of gas that cannot 
flow into the gap and indeed pushes the planet inward. Then, equations (\ref{mig_d}) and (\ref{mig_p}) can be re-written as
\begin{equation}
\tau_{mig,d}=\frac{M_d(r_p)}{\dot{M}_{mig}} = \frac{M_d(r_p)}{f_{flow}\dot{M}_{vis}}, \\
\label{eq:tmigd_flow}
\end{equation}
and
\begin{equation}
\tau_{mig,p}=\frac{M_p}{\dot{M}_{mig}} = \frac{M_p}{f_{flow}\dot{M}_{vis}},
\label{eq:tmigp_flow}
\end{equation}
where
\begin{equation}
f_{flow}= 1- \frac{\dot{M}_{cross} + \dot{M}_{p}}{\dot{M}_{vis}}.
\label{f_flow}
\end{equation}
Thus, the timescale of type II migration becomes longer if gas flow across a gap ($\dot{M_{cross}} \neq 0$) is taken into account (since $f_{flow} < 1$).

We now discuss the survival condition in detail. In the limit of $\dot{M}_{p}=0$, gas giants are saved from plunging into the central stars 
due to rapid type II migration if either $\dot{M}_{cross} \simeq \dot{M}_{vis}$ or $M_d(r_d) \la M_p$ is satisfied. As discussed above, 
the latter condition requires a fine tuning, so that the former one has a more chance. Consequently, when gas flow across a gap becomes comparable to 
the innate disk accretion rate, this process can act as an effective barrier for type II migration.

Although the recent numerical studies show that considerable amount ($\dot{M}_{cross} \sim$ 10 - 25 \% of $\dot{M}_{vis}$) of gas flows into 
the inner disk across a gap \citep[e.g.,][]{lsa99,dhk02,ld06}, the required amount of gas flow may be too high to achieve. 
One may thus tend to conclude that gas flow across a gap formed by gas giants also cannot be a solution to the problem of rapid type II migration. 
Nonetheless, we suggest that the gas flow may still have more potential to save gas giants due to the following associated processes.

First, when gas in the outer disk passes through a gap, the gas follows the horseshoe orbits around the planet. It is thus anticipated that 
the flow results in the enhancement of coronation torque. Since the inward gas flow dictates that the coronation torque transfers angular momentum from 
the gas to the planet \citep[e.g.,][]{ward91,m01,m02}, the gas flow gives rise to an additional slowing down in type II migration. 
This kind of the angular momentum transfer is very efficient and is considered as the origin of type III migration \citep{mp03}. 
It is proposed that type III migration is very fast and effective for planets with intermediate masses such as Saturn. 
Although detailed 2D hydrodynamical simulations are needed for quantitatively estimating how much amount of gas flow is 
demanded for type II migration to be halted, gas flow that amounts to a fraction of $\dot{M}_{vis}$ may be large enough.

Second, if a planet accretes gas that flows into a gap, the gas accretion can provide another slowing down 
for the planet. As shown in equation (\ref{f_flow}), $f_{flow}$ becomes small if $\dot{M}_{p} > 0$, which slows down type II migration further. 
Note that numerical simulations show that gas flow across a gap generally accompanies gas accretion onto a planet \citep[e.g.,][]{lsa99,dhk02,ld06}, 
so that the survival condition derived above ($\dot{M}_{cross} \simeq \dot{M}_{vis}$) is likely to be overestimated. 
In addition, gas accreted onto a planet mainly originates from the outer disk, rather than the inner one.
This indicates that, as the planet accretes the gas, the specific angular momentum of the planet increases. 
The increment eventually leads to outward migration. Moreover, gas accretion increases the effects of planetary inertia that also 
acts as a brake for type II migration. 

Thus, gas flow across a gap formed by a massive planet invokes the relevant processes that can serve as additional mechanisms for slowing 
down type II migration, and hence may have some potential to resolve the problem of too rapid type II migration.

As discussed above, the effects of planetary inertia are likely to be tightly coupled with gas flow around a planet and planetary growth. 
We here examine how different the fate of gas giants is by adopting different treatments. 

\citet{mab09,mabn09,rph13} assumed that type II migration starts after a growing planet satisfies the thermal condition for gap opening, that is given as 
$r_{\rm H} > H$, where $r_{\rm H}$ is the Hill radius of the planet. They nevertheless assumed that the planet keeps accreting gas of $\dot{M}_{vis}$, following 
\citet{dkh03,ld06} which showed that significant fraction of $\dot{M}_{vis}$ crosses the gap. This is essentially identical to assuming that 
\begin{equation}
\dot{M}_p =\dot{M}_{mig} =\dot{M}_{vis} (\mbox{with } \dot{M}_{cross}=0)
\end{equation}
in our formalism. Note that, with their prescription, $d\log M_p/d\log r_p = (1/M_p)(dM_p/dt)r_p/(dr_p/dt) \sim - (\dot{M}_{vis}/M_p) / (\dot{M}_{vis}/M_p) \sim -\pi$ 
(also see equation (\ref{mig_p})). Then, as a giant planet migrates, $M_p$ rapidly increases while $M_d(r_d)$ is decreased by the planetary growth.
As a result, the planet migration is halted well before the planet migrates to the proximity of the central star, except for the cases in very massive disks.
It is however obvious that the assumption cannot satisfy equation (\ref{decomp}).

In addition, even in the simulations of \citet{dkh03,ld06}, gas flow across a gap is significantly reduced after planets have grown to $\sim 10 M_{J}$, 
where $M_{J}$ is the mass of Jupiter. This trend is also supported by \citet{dll07}. Thus, it is very unlikely that planetary growth proceeds consectively 
with $\dot{M}_{p}=\dot{M}_{vis}$. Moreover, such continuous gas accretion would result in formation of 
too massive gas giants that is inconsistent with the observation. In fact, \citet{mab09,mabn09} considered relatively strong external
photoevaporation (see Section \ref{disc_pe}) to suppress the formation of too massive planets. This gives another interesting suggestion that if this 
is the case, the final stage of planetary growth that can be regulated by gas flow across a gap should be intimately linked with disk disposal mechanisms.

On the other hand, \citet{il04i,il08v} adopted a prescription that gas accretion onto planets is truncated after the thermal condition for gap opening is satisfied,
according to the results of \citet{bcl99,dll07}. In addition, they assume that type II migration starts when the viscous condition for gap opening is satisfied, 
which is usually applicable for much smaller planetary masses than those for the thermal condition. Adopting our formalism, their prescription is translated as 
\begin{equation}
\dot{M}_p=\dot{M}_{cross} (< \dot{M}_{vis})
\end{equation}
and 
\begin{equation}
\dot{M}_{mig}=\dot{M}_{vis}.
\end{equation}
This also cannot satisfy equation (\ref{decomp}). Nonetheless, the treatment of the latter may be more conservative than that of the former, since 
planetary growth is not so enhanced. As a result, type II migration is much more effective in the latter's results than in the former's ones. 
More detailed hydrodynamical simulations are needed to study the gas accretion rate onto planets after gap opening and 
the condition for the start of type II migration.

We have so far focused on type II migration of single planets. What happens for planetary migration of a pair of massive planets? 
\citet{ms01} investigated a consequence of planetary migration for a system of massive planets like Jupiter and Saturn. 
They showed that the outer light planet, that is initially located far away from the inner massive planet, 
catches up the inner one and is trapped in mean-motion resonances. Also, they demonstrated that the system migrates outward together 
after the trapping and sharing gaps formed by both planets. This outward migration arises mainly from an increased mass flow through 
the overlapping gap from the outer to the inner disk \citep[also see][]{mc07}.

Thus, it is needed to investigate in detail gas flow around a gap formed by planets for understanding type II migration more accurately.

\section{Summary \& Discussion} \label{conc}

We have analytically investigated the timescale of type II migration in gas disks that is regulated by viscous diffusion. 
We point out that theoretically predicted type II migration is so fast that gas giants migrate to the proximity of the central stars except for a very narrow window 
of timing between formation of gas giant planets and disk gas depletion. Such efficient migration is inconsistent with the semimajor axis distribution of 
extrasolar gas giants obtained by radial velocity surveys. The inertia of a planet, the presence of a MRI dead zone, and internal/external photoevaporation 
make type II migration less efficient. However, none of them saves gas giants from significant migration with the current knowledge of each process, 
if gas flow across a gap from the outer disk to the inner disk or that accreted by a planet is negligible.

We formulate type II migration and disk lifetimes based on disk accretion rates that push planets inward and remove the disk. 
With this formulation, we can consistently discuss comparison between the migration and disk depletion timescales, and 
the effects of the planetary inertia, dead zones, photoevaporation, and gas flow across a gap formed by a massive planet.

The "planet-dominated" regime where the inertia of a planet  decelerates the migration is distinguished from "disk-dominated" regime by the condition 
that planetary mass, $M_p$, is larger than the disk mass within planetary orbit, $M_d(r_p)$, where $r_p$ is the planet's orbital radius 
(see equation (\ref{crit_mass})). We have adopted a standard self-similar solution for quantifying both the local and global timescales of the disks. 
In this solution, the disk accretion rate $\dot{M}_{vis}$ is constant in the regions well inside the disk size, $r_d$. 
The local viscous timescale is given by $M_d(r_p)/\dot{M}_{vis}$, and it can be used as the timescale of disk-dominated type II migration,
whereas the disk lifetime is given by $M_d(r_d)/\dot{M}_{vis}$. The timescale of planet-dominated type II migration is given by $M_p/\dot{M}_{vis}$ 
(see equation (\ref{mig_p1})).

Since most disk masses distribute in the outer disk and it is usually expected that $r_p \ll r_d$, then $M_d(r_p) \ll M_d(r_d)$. 
Thus, we have shown that gas giants can survive against type II migration only if they are formed in the final stage of disk evolution 
in which $M_p > M_d(r_d)$ -- a fine tuning is needed for preventing the loss of gas giant planets into the host stars. 
This kind of problem in type II migration has been overlooked in the literature because the problem of type I migration is very serious. 

We have examined additional physical processes that can be present in more detailed disk models: dead zones, photoevaporation of gas, and 
gas flow across a gap formed by a type II migrator. They may be considered as potential agents for slowing down type II migration. 
When disks have dead zones, most disk masses are contained in a dead zone. As a result, the radius of an outer boundary of the dead zone must 
be effectively regarded as $r_d$, which makes $M_d(r_d)$ closer to $M_d(r_p)$. However, $M_d(r_d)$ does not become smaller than $M_d(r_p)$. 
Consequently, the migration timescale cannot be longer than the disk lifetime (except in the final disk evolution stage with $M_p > M_d(r_d)$).

Internal photoevaporation is effective in a narrow region near a gravitational radius $r_g$. If $r_g > r_p$, it decreases the disk accretion rate, and 
hence planets undergo type II migration less efficiently. This may however result in gas surface density that becomes too low to produce gas giants. 
If photoevaporation is effective at $r_g < r_p$, removal of disk gas inside $r_g$ works as a barrier for migration. Only EUV photoevaporation makes $r_g$ 
as small as the radius of gas giant formation site ($r =$1--10 AU) (Note that FUV and X photoevaporation is effective in more distant regions). 
It is nonetheless inferred currently that EUV flux is too weak to open up a gap in a disk except the final disk evolution stage. 
Thus, although the problem indeed becomes less severe, dead zones and photoevaporation of gas cannot be crucial, as individual processes,
for resolving the problem of too rapid type II migration.

We note that EUV heating is the most uncertain, because EUV is readily absorbed by the interstellar matter. Thus, it is required to study EUV flux and 
its photoevaporation both observationally and theoretically in more details for discussing the effects of photoevaporation on type II migration more accurately.

If significant fraction of disk accretion crosses a gap around the planetary orbit, the disk accretion rate to push the planet is decreased.
We have suggested that associated physical processes such as coronation torque and gas accretion onto a planet can act as additional brakes 
for rapid type II migration. Thus, the flow of gas around a gap may be potentially important for solving the problem of type II migration. 
Nonetheless, it is obvious that more detailed studies are needed for examining its effect on type II migration more seriously.

It is interesting that \citet{hp12} have recently shown that the observed mass-period relation can be well reproduced 
if multiple planet traps are incorporated in disks for trapping rapid type I migration. They considered both dead zones and photoevaporation in single disks. 
Also, they focused on planet formation beyond dead zones. As discussed above, these all act for slowing down type II migration. 
That is why their results do not suffer from the problem of type II migration significantly.\footnote{Recently Hasegawa \& Pudrtiz 
(2013, submitted to ApJ) have improved their model and investigated the statistical properties of planets formed at multiple planet traps. 
They have found that, even if planet traps are incorporated, somewhat slowed down type II migration is likely to be preferred for quantitatively reproducing 
the observations of exoplanets. This is because the "initial" distribution of planets that is generated by planet traps and can reproduce the observations 
tends to be washed out by type II migration.} Obviously, it is important to investigate the problem of type II migration in detail and examine the individual 
and combined effects of dead zones, photoevaporation of gas, and gas flow across a gap formed by missive planets.

In a subsequent paper, we will undertake the task by performing numerical simulations of viscously evolving disks and 
investigating the problem of type II migration.


\acknowledgments
The authors thank an anonymous referee for useful comments on our manuscript.
YH thank the hospitality of Tokyo Institute of Technology for hosting stimulating visits during which this work was developed. 
Y.H. is supported by EACOA Fellowship that is supported by East Asia Core Observatories Association which consists of 
the Academia Sinica Institute of Astronomy and Astrophysics, the National Astronomical Observatory of Japan, the National Astronomical 
Observatory of China, and the Korea Astronomy and Space Science Institute.  






\bibliographystyle{apj}

\bibliography{apj-jour,../../citation/adsbibliography}    

\

\end{document}